\documentclass[aps,preprint,amsmath,amssymb,superscriptaddress,nofootinbib]{revtex4}

\usepackage{graphicx}
\begin{document}

\title{Photoproduction of mixed radions at a proton-proton collider}

\author{\.{I}. \c{S}ahin}
\email[]{inancsahin@ankara.edu.tr}
 \affiliation{Department of
Physics, Faculty of Sciences, Ankara University, 
Ankara, Turkey}

\author{S. Spor}
\email[]{serdar.spor@beun.edu.tr}
 \affiliation{Department of Physics, Faculty of Arts and Sciences, Bulent Ecevit University, Zonguldak, Turkey}

\author{D. Yilmaz}
\email[]{dyilmaz@eng.ankara.edu.tr}
 \affiliation{Department of
Physics Engineering, Faculty of Engineering, Ankara University, 
Ankara, Turkey}

\author{G. Akkaya Sel\c{c}in}
\email[]{akkayag@ankara.edu.tr}  \affiliation{Department of Physics, Faculty of Arts and
Sciences, Bitlis Eren University, Bitlis, Turkey} 

\begin{abstract}
We investigate Higgs-radion mixing scenario through single radion photoproduction process $pp\to p\gamma p\to pRqX$ at the LHC where $X$ represents the remnants of one of the initial protons. We consider high luminosity values
of $L_{int}=200\;fb^{-1}$, $500\;fb^{-1}$ and $3000\;fb^{-1}$. We obtain bounds on the mixing parameter space by considering $R\to \gamma\gamma$, $R\to W^+ W^-$ and $R\to ZZ$ decay channels of the radion
as the signal. We also perform a similar analysis for a 100 TeV future proton-proton collider and compare its potential with that of LHC.

\end{abstract}


\maketitle

\section*{Introduction}

The standard model (SM) of particle physics, which describes the interaction of fundamental particles,  
has been proven to be successful in the experiments at the Large Hadron Collider
(LHC) and elsewhere. The last important confirmation of the SM is the discovery of the Higgs boson whose 
existence was verified experimentally by ATLAS and CMS Collaborations \cite{Aad:2012tfa,Chatrchyan:2012xdj}. Shortly after its discovery, Randall-Sundrum (RS) radion has been discussed in some papers 
which may be responsible for this new 125 GeV excess \cite{Cheung:2011nv,Tang:2012uw,Giardino:2012dp,Chacko:2012vm}. Indeed RS radion has very similar properties with the 
SM Higgs boson. Both are scalar bosons and couple to SM particles proportional to their mass. An important difference 
is originated from trace anomaly; the coupling of the radion to photon and gluon pair is larger than that of the Higgs. 
Although it is experimentally difficult to discern radion from Higgs boson, a detailed analysis shows that the observed 
new scalar is likely to be the SM Higgs instead of RS radion \cite{Tang:2012uw,Giardino:2012dp} \footnote{We mean the radions in the RS-1 model not in some extended version of the original model.}. 
However, similarities between SM Higgs and RS radion are remarkable and lead us to think about an interesting scenario called Higgs-radion mixing 
\cite{Csaki:2000zn,Dominici:2002jv,Cheung:2003fz,Cheung:2005pg,Chaichian:2001rq,Gunion:2003px,Toharia:2008tm,Barger:2011hu,
deSandes:2011zs,Kubota:2012in,Desai:2013pga,Cox:2013rva,Frank:2016oqi,Chakraborty:2017lxp,Frank:2012nb,Grzadkowski:2012ng,Kubota:2014mma,Angelescu:2017jyj,Merchand:2018bxu}. 
According to this scenario the Higgs boson can mix with a radion field and constitute two physical mixed states. The one of the mixed state might have
a mass of 125 GeV. Hence, the observed scalar at the LHC may not be the SM Higgs but it may be Higgs-like mixed state \cite{Kubota:2012in,Frank:2016oqi}. 
The detection of the other radion-like mixed state in a collider experiment will be an evidence for the new physics and Higgs-radion mixing.     

\begin{figure}
\includegraphics[scale=0.7]{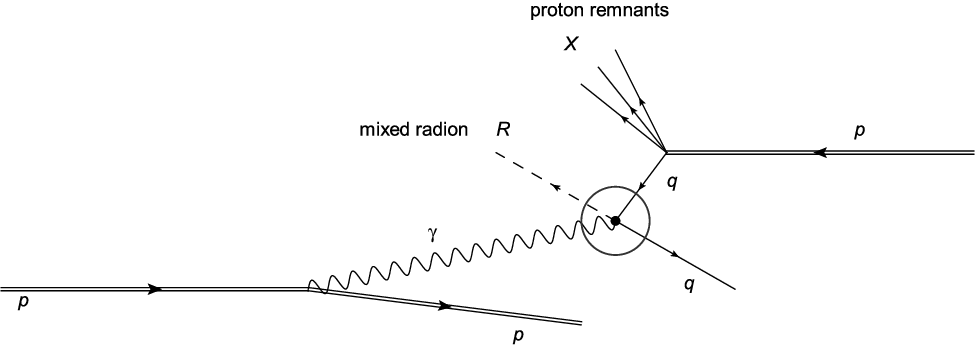}
\caption{The schematic diagram which illustrates the process $pp\to p\gamma p\to pRqX$.
\label{fig1}}
\end{figure}

In this paper we investigate Higgs-radion mixing scenario through the process $pp\to p\gamma p\to pRqX$ at the LHC and at the future proton-proton collider. Here, 
$X$ represents the remnants of one of the initial protons. These proton remnants are caused by deep $\gamma$-proton collision. 
A schematic diagram describing this reaction is given in Fig \ref{fig1}.  This process is possible at a proton-proton collider by virtue of equivalent photon
approximation (EPA) \cite{Weizsacker,Williams,budnev1975,baur2002,piotrzkowski2001}. According to EPA one or both of the incoming proton beams can emit equivalent photons having
a low virtuality $Q^2=-q_\gamma^2\approx 0$, where $q_\gamma$ is the momentum of equivalent photon. Due to this low virtuality behavior, equivalent photons can be
assumed to be real and a photon-photon or a photon-proton collision process can be studied as a subprocess in a proton-proton collision. We consider 10 independent 
subprocesses $\gamma q \to R q$, where $q=u,d,s,c,b,\bar u,\bar d,\bar s,\bar c,\bar b$ quarks. Here, $R$ represents radion-like mixed state.

Although hadron colliders are essentially designed to examine deep inelastic scattering processes, it was shown experimentally at the Fermilab Tevatron and later at the 
LHC that complementary to deep inelastic hadron collisions it is possible to study photon-photon and photon-hadron collisions via elastic photon emission in a hadron collider 
\cite{cdf1,cdf2,cdf3,Chatrchyan:2011ci,Chatrchyan:2012tv,Chatrchyan:2013foa,Khachatryan:2016mud,Aaboud:2016dkv,Hagiwara:2017fye,Cms:2018het,Aaboud:2017oiq}. Recent results from CMS and ATLAS Collaborations 
show that these photon processes at the LHC poses a considerable potential to probe new physics beyond the SM \cite{Chatrchyan:2013foa,Khachatryan:2016mud,Aaboud:2016dkv}. One 
important advantage of photon processes in comparison to deep inelastic hadron collisions is that they provide clean experimental channels which do not contain many QCD backgrounds 
and uncertainties from proton dissociation. This makes it easy to determine any possible signal which may come from new physics. The phenomenological works on new physics via photon-photon 
and photon-proton collisions at the LHC have been growing rapidly. These works embrace a wide range of models beyond the SM such as 
extra dimensions, magnetic monopoles, supersymmetry etc. It is not possible to cite all of these works, but some representative papers are given in references \cite{pheno-1,pheno-2,
pheno-3,pheno-4,pheno-5,pheno-6,pheno-7,pheno-8,pheno-9,pheno-10,pheno-11,pheno-12,pheno-13,pheno-14,pheno-15,pheno-16,pheno-17,pheno-18,pheno-19,pheno-20,
pheno-21,pheno-22,pheno-23,pheno-24,pheno-25,pheno-26,pheno-27,pheno-28,pheno-29,pheno-30,pheno-31,pheno-32,pheno-33,Das:2015toa,pheno-34,pheno-35,pheno-36,pheno-37,pheno-38,pheno-39}.

\section*{RS model of extra dimensions and the Higgs-radion mixing}

Although SM is successful in explaining fundamental particles and their interactions at the energy scale of current colliders, there are still some unanswered questions that need to be addressed. 
One such unanswered problem is the hierarchy problem which can be briefly summarized as the unexplained large energy gap between the electroweak scale and the Planck scale. Extra dimensional models 
provide a solution to the hierarchy problem. One of the popular extra dimensional models which offers a solution to the hierarchy problem is the Randall-Sundrum (RS) model of warped extra dimensions \cite{Randall:1999ee}.
This original model which assumes a small extra dimension is sometimes called RS-1 model. It shouldn't be confused with RS-2 model with an infinite extra dimension \cite{Randall:1999vf,Liu:2017gcn}. According to RS-1 model
there is one extra spatial dimension and two 3-branes located at the fixed points of the orbifold $S^1/Z_2$. If $y$ represents the extra dimensional coordinate and $r_c$ is the compactification radius then 3-branes 
are located at positions $y=0$ and $y=\pi r_c$. The 3-brane at $y=0$ is called the Planck brane or the hidden brane and the other 3-brane ($y=\pi r_c$) is called the TeV brane or the visible brane. It is assumed that 
all SM fields are confined on the TeV brane whereas the gravity can propagate into the bulk. 5-dimensional bulk is an anti-de Sitter space and has a cosmological constant $\Lambda$. Its geometry is described by the metric
\cite{Randall:1999ee}
\begin{eqnarray}
\label{RSmetric}
ds^{2}=e^{-2k|y|}\eta_{\mu\nu}dx^{\mu}dx^{\nu}-dy^{2},
\end{eqnarray}
where $k$ represents the bulk curvature. If the metric (\ref{RSmetric}) is substituted into the action and the $y$ coordinate is integrated out then the following relation between the Planck
scale ($\bar{M}_{Pl}$) and the fundamental scale ($M$) is deduced:
\begin{eqnarray}
\bar{M}^{2}_{Pl}=\frac{M^{3}}{k}(1-e^{-2kr_{c}\pi}).
\end{eqnarray}
The large mass difference between the Planck scale and the fundamental scale can be eliminated if we choose $k\approx \bar{M}_{Pl}$ and $k r_c \pi\approx 35$. Consequently, the hierarchy problem is solved. 
In the original paper of the RS-1 model \cite{Randall:1999ee}, the compactification radius is associated with the vacuum expectation value of a scalar field. However, $r_c$ is not determined by the dynamics and hence 
the value of $r_c$ is somewhat arbitrary. The metric in (\ref{RSmetric}) can be considered as a background metric and the fluctuation in the $y$-coordinate results in a scalar field called the radion. Goldberger and Wise 
proposed a mechanism for stabilizing the size of the fifth dimension \cite{Goldberger:1999uk}. They introduce a bulk scalar field propagating in the background metric with interaction terms on the branes generating a 
potential that can stabilize the radion. The mass and the vacuum expectation value of the radion can be determined from this potential. The metric in 5 dimensions can be written as \cite{Csaki:2000zn,Dominici:2002jv,Kubota:2012in}
\begin{eqnarray}
\label{Rsmetric2}
ds^{2}=e^{-2(kr_c|\phi|+F(x,\phi))}\eta_{\mu\nu}dx^{\mu}dx^{\nu}-(1+2F(x,\phi))^2r_c^2d\phi^{2}.
\end{eqnarray}
Here, $F(x,\phi)$ represents the scalar perturbation and it can be given in the following form: $F(x,\phi)=\Phi(x)R(\phi)$. In this formula, the function $R(\phi)$ is determined by demanding that
the metric in (\ref{Rsmetric2}) solves Einstein field equations. $\Phi(x)$ represents the 4-dimensional radion field which is canonically normalized. The approximate solution for the scalar perturbation
is given by 
\begin{eqnarray}
\label{perturbation}
F(x,\phi)=\frac{\Phi(x)}{\Lambda_\phi}\exp{\{2kr_c(\phi-\pi)\}}
\end{eqnarray}
where $\Lambda_\phi=\sqrt{6}M_{Pl}\exp{(-kr_c\pi)}$ \cite{Csaki:2000zn,Dominici:2002jv}.

Various extended versions of the RS-1 model were considered in the literature. In some versions of the model, not only the gravity but also some other particles are allowed to propogate in the extra dimension. Consequently, 
there are Kaluza-Klein (KK) excitations of these bulk particles; but in many extended versions of the RS-1 model these KK excitations are very heavy and their direct contribution to the processes at the LHC energies can be ignored. 
On the other hand, the mass of the radion can be smaller than the TeV scale and can be directly detected at the LHC \cite{Kubota:2012in,Kubota:2014mma,Desai:2013pga,Cox:2013rva,Frank:2016oqi}. We assume such a scenario and employ 
the same formalism used in Refs. \cite{Kubota:2012in,Frank:2016oqi}.

The radion couplings to massive gauge bosons and fermions have a similar form to that of the Higgs boson. The only difference is that the radion couplings are inversely proportional to $\Lambda_\phi$ whereas Higgs couplings are 
inversely proportional to the electroweak scale $\nu$. On the other hand, the couplings of the radion to photon and gluon receive additional contributions from tree-level couplings \cite{Csaki:2007ns} and also from trace anomalies
\cite{Csaki:2000zn}. The interaction Lagrangian for the Higgs boson and the radion to the massless gauge bosons are given by \cite{Kubota:2012in,Frank:2016oqi}
\begin{eqnarray}
\label{intlag1}
{\cal{L}}_h=&&\frac{1}{8\pi\nu}\left\{\alpha b_{EM}^h\; hF_{\alpha\beta}F^{\alpha\beta}+\alpha_s b_{QCD}^h\; hG^{(a)}_{\alpha\beta}G^{(a)\alpha\beta}\right\} \\
\label{intlag2}
{\cal{L}}_r=&&\frac{1}{4\Lambda_\phi}\left\{\left(\frac{\alpha}{2\pi}b_{EM}^r +\frac{1}{kr_c\pi}\right)\;rF_{\alpha\beta}F^{\alpha\beta}
+\left(\frac{\alpha_s}{2\pi}b_{QCD}^r +\frac{1}{kr_c\pi}\right)\;rG^{(a)}_{\alpha\beta}G^{(a)\alpha\beta}\right\}
\end{eqnarray}
where $h$ and $r$ represent Higgs and radion fields, $\alpha$ and $\alpha_s$ are the electromagnetic and strong coupling constants and 
\begin{eqnarray}
&&b_{QCD}^h=F_f ,\;\;\;\;\;\;\;\;\;\;\;\;\;\;\;\;\;\;\;\;b_{QCD}^r=7+F_f \\
\;\;\;\;\;&&b_{EM}^h=\frac{8}{3} F_f-F_W ,\;\;\;\;\;\;\;\;\;\;b_{EM}^r=-\frac{11}{3}+\frac{8}{3} F_f-F_W \\
&&F_f=\tau_f\left(1+(1-\tau_f)f(\tau_f)\right) \\
&&F_W=2+3\tau_W+3\tau_W(2-\tau_W)f(\tau_W)
\end{eqnarray}
\begin{eqnarray}
&&f(\tau)=\left\{\begin{array}{cc}
                                 \left(\arcsin\frac{1}{\sqrt {\tau}}\right)^2 \;\;\; ;\tau>1&  \\
                                -\frac{1}{4} \left(\log\frac{\eta_+}{\eta_-}-i\pi\right)^2 \;\;\; ;\tau<1&  \\
                              \end{array}\right\}\\
&&\tau=\left(\frac{2m_1}{m_2}\right)^2,\;\;\;\;\eta_\pm=1\pm\sqrt{1-\tau}.
\end{eqnarray}
Here, $m_1$ represents the mass of the particle in the loop and $m_2$ represents either the radion or the Higgs mass. We do not give the interaction Lagrangians of the radion and Higgs to other SM particles. These can be found 
in the literature. For example see Ref.\cite{Kubota:2012in} or \cite{Frank:2016oqi}. 

The mixing between the Higgs and the radion fields can be generated by the following action \cite{Csaki:2000zn,Dominici:2002jv}
\begin{eqnarray}
\label{mixingaction}
S_{\xi}=\xi\int dx^4 \sqrt{g_{TeV}}R(g_{TeV})H^\dagger H
\end{eqnarray}
where $g_{TeV}$ is the induced metric on the TeV brane, $R(g_{TeV})$ is the coresponding 4-dimensional Ricci scalar and $\xi$ is the mixing parameter. The effective action (\ref{mixingaction}) give rise to the following
Lagrangian containing bilinear fields
\begin{eqnarray}
\label{mixlagrangian}
 {\cal{L}}_{mix}=-\frac{1}{2}(1+6\gamma^2\xi)r \Box r-\frac{1}{2}m_r^2r^2-\frac{1}{2}m_h^2h^2-\frac{1}{2} h \Box h+6\xi\gamma h \Box r  
\end{eqnarray}
where $m_r$ and $m_h$ are the radion and the Higgs mass and $\gamma=\nu/\Lambda_\phi$. The kinetic part of the above Lagrangian can be diagonalized by redefining the radion and the Higgs fields. The transformation 
which diagonalizes the Lagrangian is given by \cite{Dominici:2002jv,Kubota:2012in}
\begin{eqnarray}
\label{mix}
h=dH+cR ,\;\;\;\;\;\;\;\;\;\;\;\;\;\;\;\;\;\;r=aR+bH 
\end{eqnarray}
where
\begin{eqnarray}
\label{mix2}
&&d=\cos {\theta}-\frac{6\xi\gamma}{Z}\sin {\theta},\;\;\;\;\;\;\;\;c=\sin {\theta}+\frac{6\xi\gamma}{Z}\cos {\theta} \\
&&a=\frac{\cos {\theta}}{Z}, \;\;\;\;\;\;\;\;\;\;\;\;\;\;\;\;\;\;\;\;\;\;\;\;\;b=-\frac{\sin {\theta}}{Z} \\
&&Z=\sqrt{\beta-36\xi^2\gamma^2}, \;\;\;\;\;\;\;\;\;\;\;\beta=1+6\xi\gamma^2.
\end{eqnarray}
Here, the angle $\theta$ can be solved from the following equation:
\begin{eqnarray}
 \tan{2\theta}=\frac{12\xi\gamma Z m_h^2}{m_r^2-m_h^2\left(Z^2-36\xi^2\gamma^2\right)}
\end{eqnarray}
In eqn.(\ref{mix}) the mass eigenstates are represented by $R$ and $H$. These are mixed fields containing both $h$ (SM Higgs) and $r$ (RS radion). The notation is due to the fact that when the mixing parameter converges to zero
($\xi \to0$), the fields $h$ and $r$ converge to $H$ and $R$ respectively. Therefore, $H$ refers to {\it Higgs-like} and $R$ refers to  {\it radion-like} mixed state. We will denote the masses of these mixed scalars by $m_H$ and 
$m_R$. Their values can be obtained by diagonalizing  the Lagrangian (\ref{mixlagrangian}) through the transformation (\ref{mix}) and given by
\begin{eqnarray}
\label{mass}
 m_{\pm}^2=\left[m_r^2+\beta m_h^2\pm \sqrt{\left(m_r^2+\beta m_h^2\right)^2-4m_r^2m_h^2Z^2} \right]/2Z^2
\end{eqnarray}
where $m_+=m_R$ and $m_-=m_H$ if $m_R>m_H$ and $m_+=m_H$ and $m_-=m_R$ if $m_R<m_H$. In the first case ($m_R>m_H$), the masses for $h$ and $r$ fields are given as a function of $m_+$, $m_-$ by
\begin{eqnarray}
\label{mass1}
 m_r^2=\frac{Z^2}{2}\left[m_+^2+m_-^2+ \sqrt{\left(m_+^2+m_-^2\right)^2-\frac{4m_+^2m_-^2\beta}{Z^2}} \right] \\
 \label{mass2}
 m_h^2=\frac{Z^2}{2\beta}\left[m_+^2+m_-^2- \sqrt{\left(m_+^2+m_-^2\right)^2-\frac{4m_+^2m_-^2\beta}{Z^2}} \right]
\end{eqnarray}
and in the second case ($m_R<m_H$) they are given by
\begin{eqnarray}
\label{mass3}
 m_r^2=\frac{Z^2}{2}\left[m_+^2+m_-^2- \sqrt{\left(m_+^2+m_-^2\right)^2-\frac{4m_+^2m_-^2\beta}{Z^2}} \right] \\
 \label{mass4}
 m_h^2=\frac{Z^2}{2\beta}\left[m_+^2+m_-^2+ \sqrt{\left(m_+^2+m_-^2\right)^2-\frac{4m_+^2m_-^2\beta}{Z^2}} \right].
\end{eqnarray}
We observe from (\ref{mass1})-(\ref{mass4}) that the term inside the square root must be positive for physical masses. Therefore we have a condition which must be satisfied for $m_R$ and $m_H$:
\begin{eqnarray}
\label{allowedregion}
\left(m_R^2+m_H^2\right)^2-\frac{4m_R^2m_H^2\beta}{Z^2}>0.
\end{eqnarray}
The interaction Lagrangians for mixed-radion and mixed-Higgs to massless gauge bosons can be obtained from Lagrangians (\ref{intlag1}) and (\ref{intlag2})
by means of the transformation (\ref{mix}). These can be written as
\begin{eqnarray}
 {\cal{L}}_H=\frac{1}{8\pi\nu}\left\{\left[d\;\alpha\; b_{EM}^h+b\gamma\left(\frac{2}{kr_c}+\alpha b_{EM}^r\right) \right] HF_{\alpha\beta}F^{\alpha\beta}
\right.\nonumber \\ \left. +\left[ d\;\alpha_s\; b_{QCD}^h+b\gamma\left(\frac{2}{kr_c}+\alpha_s b_{QCD}^r\right) \right] HG^{(a)}_{\alpha\beta}G^{(a)\alpha\beta}\right\} 
\end{eqnarray}
\begin{eqnarray}
 {\cal{L}}_R=\frac{1}{8\pi\nu}\left\{\left[c\;\alpha\; b_{EM}^h+a\gamma\left(\frac{2}{kr_c}+\alpha b_{EM}^r\right) \right] RF_{\alpha\beta}F^{\alpha\beta}
\right.\nonumber \\ \left. +\left[ c\;\alpha_s\; b_{QCD}^h+a\gamma\left(\frac{2}{kr_c}+\alpha_s b_{QCD}^r\right) \right] RG^{(a)}_{\alpha\beta}G^{(a)\alpha\beta}\right\}. 
\end{eqnarray}
Eventually, the Higgs-radion mixing scenario is described by 4 parameters: $\xi$, $\Lambda_\phi$, $m_R$ and $m_H$. However, if we assume that the observed scalar at the LHC 
is not the SM Higgs $h$ but it is Higgs-like mixed state $H$, then we should take $m_H=125$ GeV and thus 3 independent parameters remain.

For the aim of completeness, we give the lagrangian that describes $\gamma Z R$ interaction. This lagrangian receives contributions from fermions and $W$ bosons circulating in the loop. It is given by 
\cite{Bhattacharya:2014wha,Gunion:1987ke,Farina:2015dua}
\begin{eqnarray}
 {\cal{L}}_{R\gamma Z}=\frac{\alpha}{4\pi \sin \theta_W}\left[A^h \frac{c}{\nu}+A^r \frac{a}{\Lambda_\phi} \right] RF_{\alpha\beta}Z^{\alpha\beta}.
\end{eqnarray}
where
\begin{eqnarray}
 A^{r,h}=A^{r,h}_1(x_W,\lambda_W)+\sum_f \frac{N_c Q_f v_f}{\cos \theta_W}A^{r,h}_{1/2}(x_f,\lambda_f).
\end{eqnarray}
The explicit expressions for the functions $A^{r,h}_1(x_W,\lambda_W)$ and $A^{r,h}_{1/2}(x_f,\lambda_f)$ and also other necessary definitions can be found in Refs.\cite{Bhattacharya:2014wha,Gunion:1987ke} 
(see also Ref.\cite{Farina:2015dua}).

\section*{The cross section of $pp\to p\gamma p\to pRqX$ and Numerical Results}

The cross section for single mixed-radion production $pp\to p\gamma p\to pRqX$ can be obtained by integrating the cross section
for the subprocesses $\gamma q\to Rq$ over the equivalent photon and quark distributions $\frac{dN_\gamma}{dx_1}$ and $\frac{dN_q}{dx_2}$:
\begin{eqnarray}
\label{mainprocess}
 \sigma\left(pp\to p \gamma p\to p R q X\right)=\sum_q\int_{{x_1}_{min}}^{{x_1}_{max}} {dx_1 }\int_{0}^{1}
{dx_2}\left(\frac{dN_\gamma}{dx_1}\right)\left(\frac{dN_q}{dx_2}\right)
\hat{\sigma}_{\gamma q\to R q}(\hat s).
\end{eqnarray}
where the sum is performed over $q=u,d,s,c,b,\bar u,\bar d,\bar s,\bar c,\bar b$ quarks. Therefore, 10 independent subprocesses are assumed to contribute to the main process. In the integral 
(\ref{mainprocess}), $x_1$ represents the energy ratio between the equivalent photon and the initial proton and $x_2$ is the momentum fraction of the proton's momentum carried by the quark. The equivalent
photon distribution function is given by the following formula \cite{Weizsacker,Williams,budnev1975,baur2002,piotrzkowski2001,pheno-4}
\begin{eqnarray}
\label{spectrum} \frac{dN_\gamma}{dx_1}=\frac{\alpha}{\pi
x_1}
\left(1-x_1\right)\left[\varphi\left(\frac{Q^{2}_{max}}{Q^{2}_0}\right)
-\varphi\left(\frac{Q^{2}_{min}}{Q^{2}_0}\right)\right]
\end{eqnarray}
where,  $Q^{2}_{0}=0.71 \mbox{GeV}^{2}$, $Q^{2}_{min}=\frac{m^{2}_{p}x_1^2}{(1-x_1)}$, $Q^{2}_{max}=2\;GeV^2$ and the $\varphi$ is defined by
\begin{eqnarray}
\varphi(x)=(1+ay)\left[-ln(1+\frac{1}{x})+\sum_{k=1}^3\frac{1}{k(1+x)^k}\right]+\frac{y(1-b)}{4x(1+x)^3}\nonumber\\
+c\left(1+\frac{y}{4}\right)\left[ln\left(\frac{1-b+x}{1+x}\right)+\sum_{k=1}^3\frac{b^k}{k(1+x)^k}\right]
\end{eqnarray}
\begin{eqnarray}
y=\frac{x_1^2}{(1-x_1)},\;\;\;\;\;\;a=\frac{1+\mu^{2}_{p}}{4}+\frac{4m^2_{p}}{Q^{2}_0}\approx7.16 \nonumber\\
b=1-\frac{4m^2_{p}}{Q^{2}_0}\approx-3.96,\;\;\;\;\;\;c=\frac{\mu^{2}_{p}-1}{b^4}\approx0.028
\end{eqnarray}
In the above formula, $m_{p}$ and $\mu_{p}$ represent the mass and the magnetic moment of the proton. In our calculations, we evaluate the quark distributions numerically by using a code MSTW2008 \cite{Martin:2009iq}.
The upper and lower limits of the $x_1$ parameter depend on the momentum fraction loss of the photon-emitting proton. After elastic photon emission, the proton deviates slightly from its trajectory along
the beam path. This deviation is related to the momentum which is transferred to the photon and described by the parameter $\xi$.\footnote{One should be careful not to confuse it with the mixing parameter.} 
The $\xi$ parameter can be given by the formula $\xi\equiv(|\vec{p}|-|\vec{p}^{\,\,\prime}|)/|\vec{p}|$ where $\vec p$ represents the initial proton's momentum and $\vec{p}^{\,\,\prime}$ represents 
forward proton's momentum after elastic photon emission. At high energies $(E\gg m_p)$ it is a good approximation to assume $\xi$ and $x_1$ are equal, $\xi\approx x_1$. The range of $\xi$ parameter is specified by 
the forward detector acceptance. Forward detectors are capable to detect scattered protons after elastic photon emission. The detection of such scattered protons by forward detectors is used to reconstruct the collision
kinematics and consequently $\gamma \gamma$ and $\gamma$-proton processes in a proton-proton collider can be identified. The LHC is equipped with such forward detectors. The ATLAS Forward Proton Detector \cite{Adamczyk:2015cjy}
and the CMS-TOTEM Precision Proton Spectrometer \cite{Albrow:2014lrm,Albrow:2013yka} are the forward detectors which can serve this purpose. Indeed the process $pp\to p\gamma \gamma p\to p \ell \bar \ell p$ is observed 
at the LHC using the proton tagging method by CMS-TOTEM Precision Proton Spectrometer \cite{Cms:2018het}. These forward detectors cover an acceptance range of $0.015<\xi<0.15$ \cite{pheno-33,pheno-36}. 
Therefore during calculations we set ${x_1}_{max}=0.15$ and ${x_1}_{min}=0.015$.

\begin{figure}
\includegraphics[scale=1]{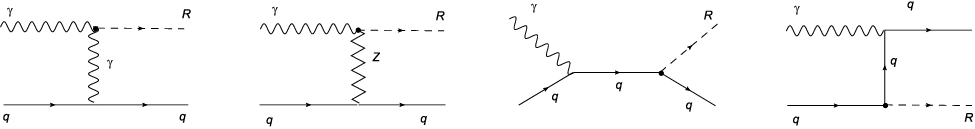}
\caption{The Feynman diagrams for the subprocess $\gamma q \to R q$.
\label{fig2}}
\end{figure}

The subprocess $\gamma q\to Rq$ is described by four tree-level Feynman diagrams (see Fig.\ref{fig2}). The biggest contribution comes from t-channel diagram that contains the $\gamma\gamma R$ vertex. As we have mentioned before,
this vertex receives additional contributions from tree-level couplings and also from trace anomalies. The quark-radion couplings also have tree-level contributions. But these are proportional to the quark mass and give minor 
contributions for light quarks. Moreover, the quark distributions for heavy quarks ($b$ and $t$) in the proton are considerably small compared with valance quark distributions ($t$ quark distribution can be safely neglected.).

\begin{figure}
\includegraphics[width=10cm,height=10cm,scale=1]{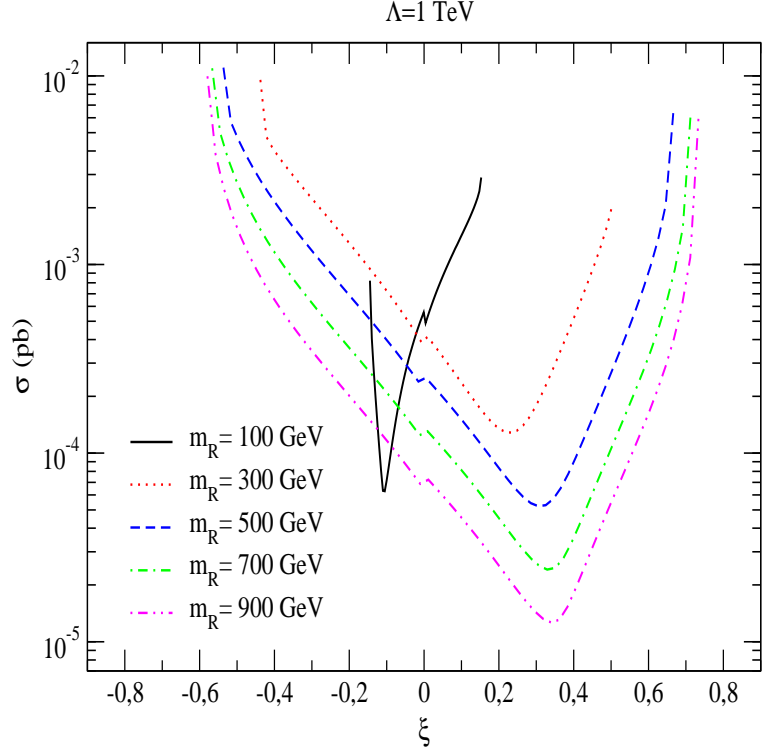}
\caption{The total cross section of the process $pp\to p\gamma p\to pRqX$ as a function of the mixing parameter $\xi$ for various values of the mixed radion mass $m_R$ stated on the figure.
The center-of-mass energy of the proton-proton system is taken to be $\sqrt{s}=14$ TeV and $\Lambda_\phi=1$ TeV. 
\label{fig3}}
\end{figure}

\begin{figure}
\includegraphics[width=10cm,height=10cm,scale=1]{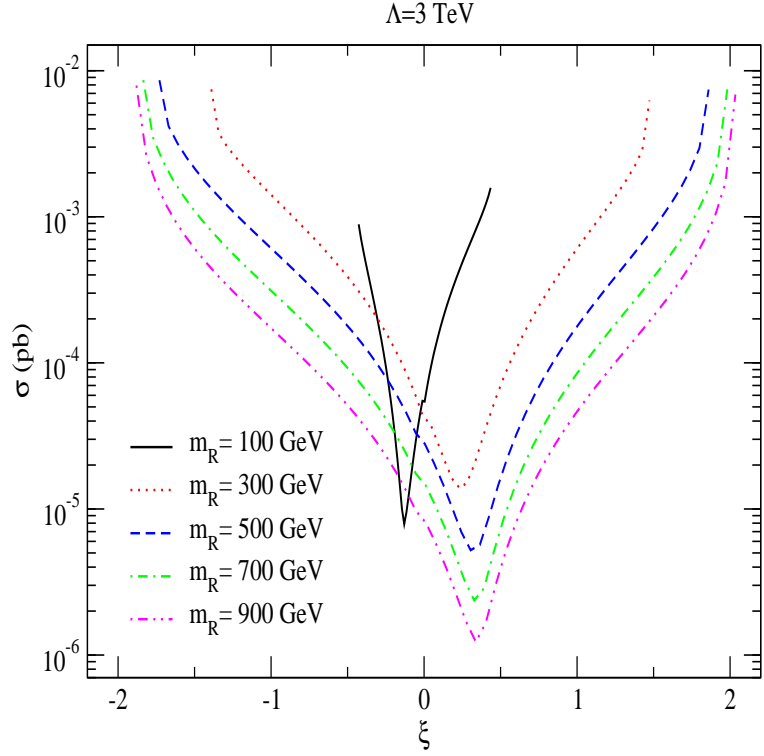}
\caption{The total cross section of the process $pp\to p\gamma p\to pRqX$ as a function of the mixing parameter $\xi$ for various values of the mixed radion mass $m_R$ stated on the figure.
The center-of-mass energy of the proton-proton system is taken to be $\sqrt{s}=14$ TeV and $\Lambda_\phi=3$ TeV. 
\label{fig4}}
\end{figure}

\begin{figure}
\includegraphics[width=10cm,height=10cm,scale=1]{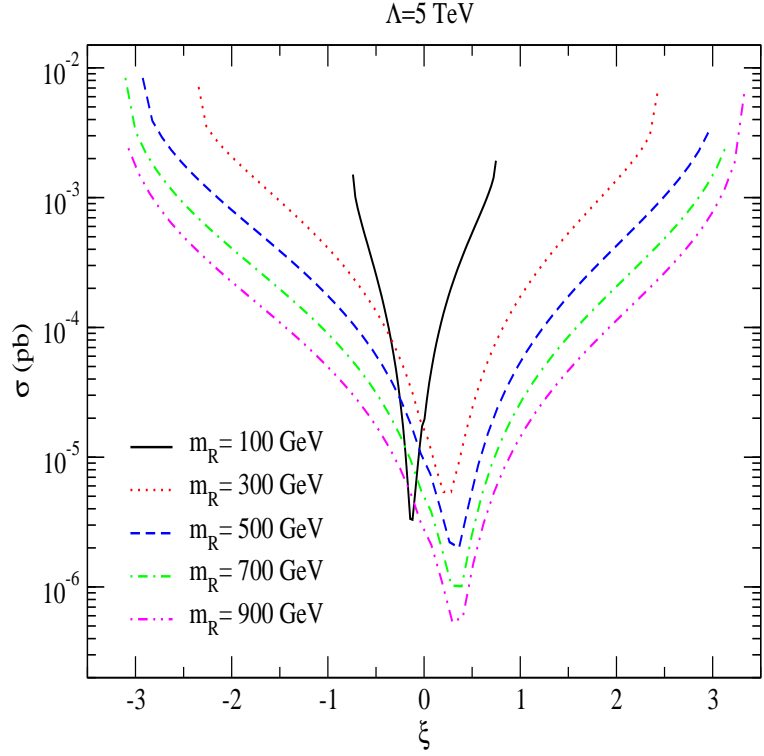}
\caption{The total cross section of the process $pp\to p\gamma p\to pRqX$ as a function of the mixing parameter $\xi$ for various values of the mixed radion mass $m_R$ stated on the figure.
The center-of-mass energy of the proton-proton system is taken to be $\sqrt{s}=14$ TeV and $\Lambda_\phi=5$ TeV. 
\label{fig5}}
\end{figure}

In Figs.\ref{fig3}-\ref{fig5} we plot the total cross section of the process $pp\to p\gamma p\to pRqX$ as a function of the mixing parameter $\xi$ for various values of the mixed radion mass $m_R$ and the scale $\Lambda_\phi$. 
We observe from these figures that the range of the $\xi$ increases when $m_R$ and $\Lambda_\phi$ increases. This behavior originates from Eq.(\ref{allowedregion}) which gives us the theoretically allowed region for $\xi$. We 
also observe from the figures that for a fixed value of the $\xi$ parameter, the cross section increases as $\Lambda_\phi$ decreases. The minima of the plots deviate slightly as a function of $m_R$ for large mass
values greater than the mixed Higgs mass. However, the curve for $m_R = 100$ GeV is very different from other curves. This behavior originates 
from the term $\Delta\equiv \left(m_+^2+m_-^2\right)^2-\frac{4m_+^2m_-^2\beta}{Z^2}$ in the square root in Eqs.(\ref{mass1})-(\ref{mass4}). 
$\Delta$ exhibits a drastic change at $m_+\approx m_-$. The masses $m_r$ and $m_h$ and the $\theta$ parameter (see Eq.(19)) change 
slightly as a function of $m_R$ in the interval which is much more greater than the mixed Higgs mass. On the other hand, when $m_R$
becomes closer to $m_H=125$ GeV, $m_r$, $m_h$ and $\theta$ change more rapidly. This explains the reason why the 
cross section curve for $m_R = 100$ GeV is very different from the curves for $m_R = 300-900$ GeV.

The mixed radions produced via this process can be detected by the central detectors through their decay products. Therefore, the invariant mass measurement of the radion decay products is crucial to determine the radion mass and
discern the process $pp\to p\gamma p\to pRqX$ from some possible SM background processes. We analyze the process by considering three different decay channel of the mixed radion:  $R\to \gamma\gamma$, $R\to W^+ W^-$ and $R\to ZZ$.  
In the first case, the number of events is given by  $N=S\times L_{int}\times \sigma(pp\to p\gamma p\to pRqX)\times BR(R\to \gamma\gamma)$, where $BR(R\to \gamma\gamma)$ is the branching ratio, $L_{int}$ is the integrated luminosity 
and $S$ is the survival probability factor which is taken to be $0.7$. The invariant mass of the final photon pair can be evaluated using the photon energies and the angle between the photons measured in the electromagnetic calorimeter. 
The uncertainty associated with photon detection is generally lower than the uncertainties associated with the detection of other particles (quarks, gluons or leptons as well). Therefore, we assume that the radions, if they exist, 
can be detected and discerned from the SM signals by performing an invariant mass measurement on the final photon pair. Hence, the number of observed events in the SM prediction is assumed to be zero. Therefore, during statistical
analysis for $R\to \gamma\gamma$ decay channel, we employ the Poisson distribution formula to constrain the free model parameters $\xi$, $\Lambda_\phi$ and $m_R$. In the Poisson distribution formula, the upper limit of the number 
of events $N_{up}$ gives $3$ events at 95\% C.L \cite{booklet}. The upper limit of number of events can be converted to the limits of model parameters through the formula $N(\xi,\Lambda_\phi,m_R,L_{int})=N_{up}=3$. Therefore, 
the restricted region in the parameter space corresponds to the values of the model parameters which satisfy $N\geq3$. However we should note that our analysis for $R \to \gamma \gamma$ is severely over-simplified. Since we have ignored 
all background processes, our analysis gives a rough estimate about the constraints on the model parameters. Hence, it is better to call these constraints 3 or more signal event constraints rather than 95\% C.L. constraints. During numerical 
calculations, we also impose a pseudorapidity cut of $|\eta|<2.5$ for all final state particles.
In Fig.\ref{fig6} we show the restricted regions in the $\xi-m_R$ parameter plane for the integrated luminosites of $L_{int}=200\;fb^{-1}$, $500\;fb^{-1}$ and $3000\;fb^{-1}$. Three panels from left to right show the restricted regions
for three different values of the energy scale $\Lambda_\phi$= $1\;\text{TeV}$, $3\;\text{TeV}$ and $5\;\text{TeV}$. 

The statistical analysis in the case of $R\to W^+ W^-$ and $R\to ZZ$ decay channels is more complicated compared with $R\to \gamma\gamma$ case. In order to discern the radions experimentally we should determine the invariant mass of 
the final $WW$ and $ZZ$ pairs with some precision. However, the uncertainties associated with $W$ and $Z$ boson detection are considerably larger than the uncertainties associated with photon detection. We assume that the invariant mass of the final particle pairs can be determined with a 20 GeV inaccuracy ,i.e., $m_R-10 \;\text{GeV}<m_{inv}<m_R+10 \;\text{GeV}$. 
There are some SM processes which give the same final states. These SM backgrounds cannot be eliminated even if the above invariant mass cut is imposed. For instance, there are 26 SM subprocesses of the form $\gamma q\to W^+ W^-q^{\prime}$
which contribute to the process $pp\to p\gamma p\to p W^+ W^-qX$. The sum of all these SM contributions gives an integrated cross section of  $1,2\times 10^{-3}$ pb when we impose an invariant mass cut of $490 \;\text{GeV}<m_{inv}<510 \;\text{GeV}$ on the final $W^+ W^-$ pair. 
This makes approximately 100 events for $L_{int}=200\;fb^{-1}$. Therefore, in the case of $R\to W^+ W^-$ and $R\to ZZ$ decay channels we employ $\chi^2$ analysis and assume that SM cross section is equal to the sum of the cross sections for 
SM backgrounds with the appropriate cuts imposed. The $\chi^2$ function is defined by
\begin{eqnarray}
\chi^{2}=\left(\frac{N_{AN}-N_{SM}}{N_{SM} \,\, \delta}\right)^{2}
\end{eqnarray}
where, $N_{AN}$ is the number of events containing both new physics
and SM contributions, $N_{SM}$ is the number of events expected in
the SM and $\delta=\frac{1}{\sqrt{N_{SM}}}$ is the statistical
error. To be precise for $R\to W^+ W^-$ case, $N_{SM}=S\times L_{int}\times \sigma_{background}(pp\to p\gamma p\to pW^+W^-qX)\times Br^2$ and $N_{AN}=S\times L_{int}\times \sigma(pp\to p\gamma p\to pRqX)\times BR(R\to W^+ W^-)\times
Br^2+ N_{SM}$, where $BR(R\to W^+ W^-)$ is the branching ratio for the $R\to W^+ W^-$ decay and $Br$ is the branching ratio of the $W$ boson to hadrons. For $R\to ZZ$ case, the number of events can be obtained in a similar manner but
we consider the branching ratios for both $Z\to \text{hadrons}$ and $Z\to \text{leptons}$. We use the following values for the branching ratios: $Br(W\to \text{hadrons})=0.674$, $Br(Z\to \text{hadrons})=0.699$ and 
$Br(Z\to \text {leptons}(e^-e^+,\mu^-\mu^+))=0.067$. The background contributions have been calculated by using CalcHEP 3.6.20 \cite{Belyaev:2012qa}.In Figs.\ref{fig7}-\ref{fig9} we present 95\% C.L. restricted regions in the 
$\xi-m_R$ parameter plane obtained from two parameter $\chi^2$ test for the processes $pp\to p\gamma p\to pRqX\to p(W^+W^-,ZZ)qX$. As in the $R\to \gamma \gamma$ case, we consider $L_{int}=200\;fb^{-1}$, $500\;fb^{-1}$ and $3000\;fb^{-1}$
integrated luminosity values and $\Lambda_\phi$= $1\;\text{TeV}$, $3\;\text{TeV}$ and $5\;\text{TeV}$ values for the energy scale. We observe from the figures that the most stringent bounds are obtained for $R\to ZZ$; $Z\to hadrons$ 
decay channels as expected. On the other hand, the restricted areas are almost same in $R\to W^+ W^-$; $W^\pm\to hadrons$ 
and $R\to ZZ$; $Z\to leptons$ decay channels. Such a result may seem contradictory at first, since the branching ratio for $W^\pm\to hadrons$ is approximately 10 times bigger than branching ratio for $Z\to leptons$. We can understand 
this result if we take account of the size of the background processes. The cross sections of the background processes for $W^+W^-$ final state are considerably larger than those for $ZZ$ final state. For instance, at the LHC
energy the sum of all background contributions gives an integrated cross section of  $1,2\times 10^{-3}$ pb when we impose an invariant mass cut of $490 \;\text{GeV}<m_{inv}<510 \;\text{GeV}$ on the final $W^+ W^-$ pair.
The corresponding background contribution for $ZZ$ final state is only $0,87\times 10^{-6}$ pb. Therefore, although the cross section for $\gamma q \to (R \to W^+ W^- \to \text{hadrons}) q$ is considerably larger than
the cross section for $\gamma q \to (R \to ZZ \to 4\; \text{leptons}) q$, sensitivity bounds are spoiled in the former process due to large background contributions.

The physics potential of a future circular collider (FCC) has been discussed by the physics community and the interest in the subject is growing rapidly \cite{Arkani-Hamed:2015vfh,Mangano:2016jyj}. Such a very high energetic 
machine is expected to have a great potential to probe the new physics. For the purpose of comparison, we have performed a similar analysis and obtain the sensitivity bounds for a proton-proton collider with the center-of-mass 
energy of 100 TeV. During calculations, we assume an integrated luminosity of $L_{int}=3000\;fb^{-1}$ and forward detector acceptance range of $0.015<\xi<0.15$ (same acceptance range with LHC). The bounds on the $\xi-m_R$ parameter 
plane are given in Fig.\ref{fig10} for $R\to \gamma \gamma$ decay channel and in Figs.\ref{fig11}-\ref{fig13} for $R\to W^+ W^-$ and $R\to ZZ$ decay channels. We plot both FCC ($\sqrt s=100$TeV) and LHC ($\sqrt s=14$TeV) bounds on
the same graph in order to provide convenience in comparison. 

\section*{Conclusions}

\begin{figure}
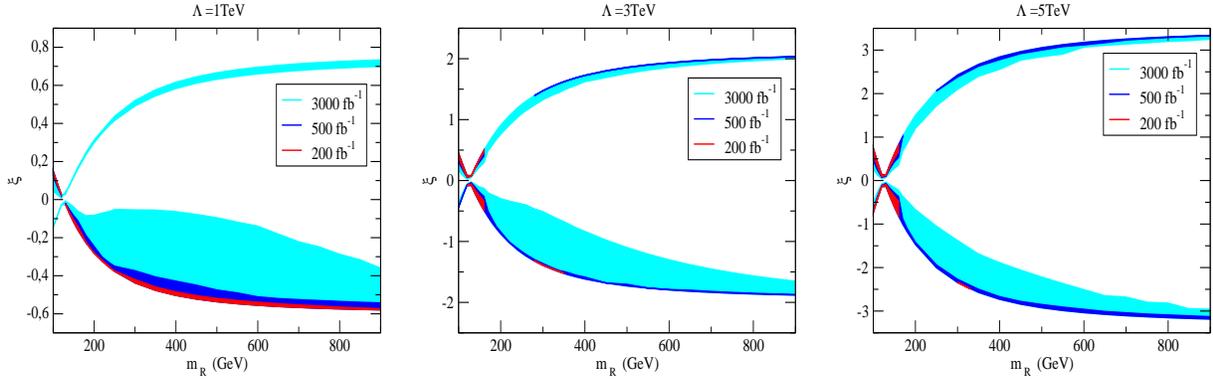
%
    \centering
\includegraphics[width=5cm,height=5cm,scale=1]{fig6.eps}\hspace{1em}
\includegraphics[width=5cm,height=5cm,scale=1]{fig7.eps}\hspace{1em}
\includegraphics[width=5cm,height=5cm,scale=1]{fig8.eps}\hspace{1em}
\caption{$N\geq3$ restricted regions in the $\xi-m_R$ parameter plane for the integrated luminosites stated on the figures. $R\to \gamma\gamma$ decay channel of
the mixed radion is considered as the signal. The center-of-mass energy of the proton-proton system is taken to be $\sqrt{s}=14$ TeV. The left, middle and right panels show
restricted regions for $\Lambda_\phi=1$ TeV, $\Lambda_\phi=3$ TeV and $\Lambda_\phi=5$ TeV respectively. 
\label{fig6}}
\end{figure}

Complementary to deep inelastic proton-proton collisions photon-photon and photon-proton collisions via elastic photon emission can be studied in a proton collider. We investigate the potential of single radion photoproduction
process $pp\to p\gamma p\to pRqX$ to probe new physics which may originate from Higgs-radion mixing. We consider high luminosity values $L_{int}=200\;fb^{-1}$, $500\;fb^{-1}$ and $3000\;fb^{-1}$ of the LHC with $\sqrt s=14$TeV.
Even though the radions cannot be detected directly by the central detectors, their existence can be inferred from their decay products. We consider $R\to \gamma \gamma$, $R\to W^+ W^-$ and $R\to ZZ$ decay channels of the radion 
as the signal. As we have shown from Fig.\ref{fig6} the restricted region obtained for $R\to \gamma \gamma$ decay channel very depends on the luminosity. For instance, at $\Lambda_\phi=3$ TeV and $m_R=400$ GeV, the restricted $\xi$ interval is 
enlarged by approximately a factor of 12 when luminosity increases from $500\;fb^{-1}$ to $3000\;fb^{-1}$. It is evident from Fig.\ref{fig6} that the restricted region almost vanishes for luminosites $L_{int}<200\;fb^{-1}$. On the other
hand, we observe from Figs.\ref{fig7}-\ref{fig9} that luminosity has a relatively minor effect on the sensitivity bounds for $R\to W^+ W^-$ and $R\to ZZ$ cases.
The experimental constraints from LHC data have been summarized in Ref.\cite{Frank:2016oqi} (see Fig.4 of \cite{Frank:2016oqi}). It was shown that the negative $\xi$ region is ruled out by the LHC data for $m_R>125$ GeV. When
we compare our bounds with the experimental bounds presented in Ref.\cite{Frank:2016oqi}, we see that our analysis for $R\to ZZ$; $Z\to hadrons$ decay channels with an integrated luminosity of $L_{int}=3000\;fb^{-1}$ and 
$\Lambda_\phi=3$ and $5$ TeV exclude almost all the allowed region above $\xi=0.5$. On the other hand, in the case of $R\to W^+ W^-$; $W^\pm\to hadrons$ and $R\to ZZ$; $Z\to leptons$ decay channels, our analysis for $\Lambda_\phi=3$ 
and $5$ TeV and $L_{int}=3000\;fb^{-1}$ exclude approximately half of the allowed region above $\xi=0.5$. 

For the purpose of comparison, we have also obtained the bounds on the $\xi-m_R$ parameter plane for a future proton-proton collider with $\sqrt s=100$TeV. We see from Figs.\ref{fig10} that when we consider $R\to \gamma \gamma$ decay
channel, FCC probes mixed radions with a far better sensitivity than the LHC. This feature is more prominent for large values of the energy scale $\Lambda_\phi$ and the mass $m_R$. On the other hand, we see from 
Figs.\ref{fig11}-\ref{fig13} that in the $R\to W^+ W^-$ and $R\to ZZ$ cases FCC provide a slight improvement in the sensitivity bounds with respect to LHC for $\Lambda_\phi=3$ and 5 TeV.  



\newpage

\clearpage

\clearpage

\begin{figure}%
    \centering
\includegraphics[width=5cm,height=5cm,scale=1]{fig12.eps}\hspace{1em}%
\includegraphics[width=5cm,height=5cm,scale=1]{fig13.eps}\hspace{1em}%
\includegraphics[width=5cm,height=5cm,scale=1]{fig14.eps}\hspace{1em}
\caption{95\% C.L. restricted regions in the $\xi-m_R$ parameter plane for the integrated luminosites stated on the figures.
In the left panel we consider $R\to W^+ W^-$; $W^\pm\to hadrons$ decay channels of the mixed radion. In the middle and right panels
we consider $R\to ZZ$; $Z\to leptons$ and $R\to ZZ$; $Z\to hadrons$ decay channels respectively.  The center-of-mass energy of the proton-proton 
system is taken to be $\sqrt{s}=14$ TeV and $\Lambda_\phi=1$ TeV.
\label{fig7}}
\end{figure}

\begin{figure}%
    \centering
\includegraphics[width=5cm,height=5cm,scale=1]{fig15.eps}\hspace{1em}%
\includegraphics[width=5cm,height=5cm,scale=1]{fig16.eps}\hspace{1em}%
\includegraphics[width=5cm,height=5cm,scale=1]{fig17.eps}\hspace{1em}
\caption{The same as FIG. \ref{fig7} but for $\Lambda_\phi=3$ TeV. 
\label{fig8}}
\end{figure}

\begin{figure}%
    \centering
\includegraphics[width=5cm,height=5cm,scale=1]{fig18.eps}\hspace{1em}%
\includegraphics[width=5cm,height=5cm,scale=1]{fig19.eps}\hspace{1em}%
\includegraphics[width=5cm,height=5cm,scale=1]{fig20.eps}\hspace{1em}
\caption{The same as FIG. \ref{fig7} but for $\Lambda_\phi=5$ TeV. 
\label{fig9}}
\end{figure}

\begin{figure}
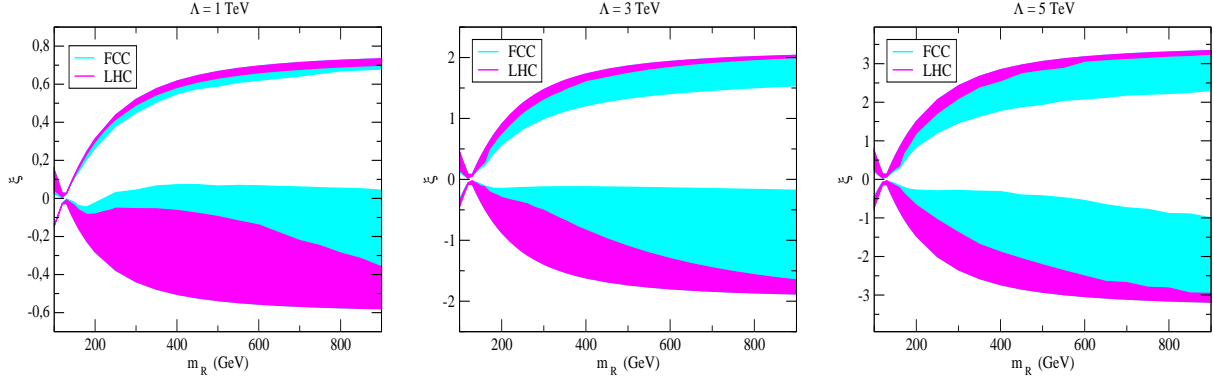
%
    \centering
\includegraphics[width=5cm,height=5cm,scale=1]{fig9.eps}\hspace{1em}
\includegraphics[width=5cm,height=5cm,scale=1]{fig10.eps}\hspace{1em}
\includegraphics[width=5cm,height=5cm,scale=1]{fig11.eps}\hspace{1em}
\caption{95\% C.L. restricted regions in the $\xi-m_R$ parameter plane for FCC ($\sqrt{s}=100$ TeV) and LHC ($\sqrt{s}=14$ TeV). $R\to \gamma\gamma$ decay channel of
the mixed radion is considered as the signal. The integrated luminosity is taken to be $L_{int}=3000\;fb^{-1}$. The left, middle and right panels show
restricted regions for $\Lambda_\phi=1$ TeV, $\Lambda_\phi=3$ TeV and $\Lambda_\phi=5$ TeV respectively. 
\label{fig10}}
\end{figure}

\begin{figure}
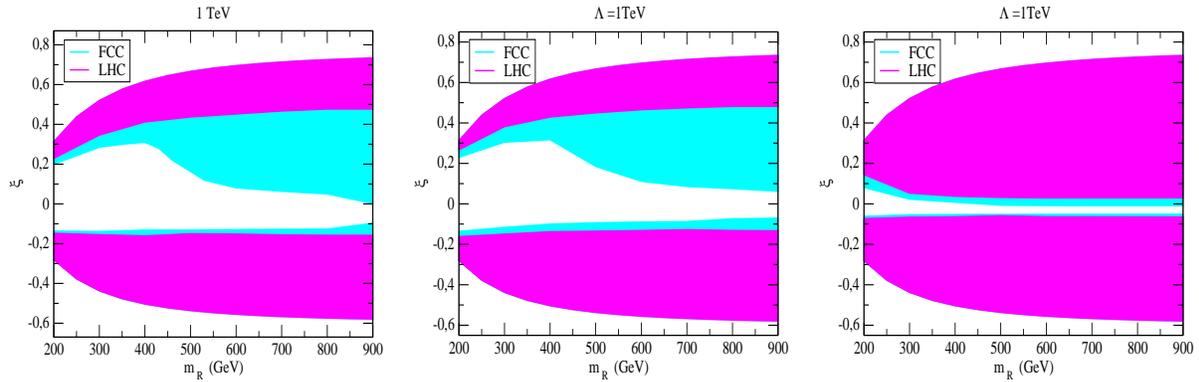
%
    \centering
\includegraphics[width=5cm,height=5cm,scale=1]{fig21.eps}\hspace{1em}%
\includegraphics[width=5cm,height=5cm,scale=1]{fig22.eps}\hspace{1em}%
\includegraphics[width=5cm,height=5cm,scale=1]{fig23.eps}\hspace{1em}
\caption{95\% C.L. restricted regions in the $\xi-m_R$ parameter plane for FCC ($\sqrt{s}=100$ TeV) and LHC ($\sqrt{s}=14$ TeV).
In the left panel we consider $R\to W^+ W^-$; $W^\pm\to hadrons$ decay channels of the mixed radion. In the middle and right panels
we consider $R\to ZZ$; $Z\to leptons$ and $R\to ZZ$; $Z\to hadrons$ decay channels respectively. The integrated luminosity is taken to be $L_{int}=3000\;fb^{-1}$
and $\Lambda_\phi=1$ TeV. 
\label{fig11}}
\end{figure}

\begin{figure}%
    \centering
\includegraphics[width=5cm,height=5cm,scale=1]{fig24.eps}\hspace{1em}%
\includegraphics[width=5cm,height=5cm,scale=1]{fig25.eps}\hspace{1em}%
\includegraphics[width=5cm,height=5cm,scale=1]{fig26.eps}\hspace{1em}
\caption{The same as FIG. \ref{fig11} but for $\Lambda_\phi=3$ TeV. 
\label{fig12}}
\end{figure}

\begin{figure}%
    \centering
\includegraphics[width=5cm,height=5cm,scale=1]{fig27.eps}\hspace{1em}%
\includegraphics[width=5cm,height=5cm,scale=1]{fig28.eps}\hspace{1em}%
\includegraphics[width=5cm,height=5cm,scale=1]{fig29.eps}\hspace{1em}
\caption{The same as FIG. \ref{fig11} but for $\Lambda_\phi=5$ TeV. 
\label{fig13}}
\end{figure}

\end{document}